\documentclass[superscriptaddress,floatfix,prl,twocolumn,amsmath,amssymb]{revtex4-1}

\usepackage{physics}
\usepackage{graphicx}
\usepackage{dcolumn}
\usepackage{bm}
\usepackage{amssymb}
\usepackage{natbib}
\usepackage{amsmath}
\usepackage{mathtools}
\usepackage{color}
\usepackage{datetime}
\usepackage{footnote}
\usepackage{bbold}
\usepackage[normalem]{ulem}
\usepackage{dcolumn}
\usepackage{ragged2e}
\usepackage{xfrac}
\usepackage{float}
\usepackage[caption=false]{subfig}
\usepackage{sidecap}
\usepackage{setspace}
\usepackage{multirow}
\usepackage{xcolor}
\usepackage{tikz}
\usepackage{tikz-cd}
\usepackage{booktabs}
\usepackage{flushend}

\usepackage[ colorlinks = true,
             linkcolor = blue,
             urlcolor  = blue,
             citecolor = red,
             anchorcolor = green,
]{hyperref}

\graphicspath{{./Figures/}}

\begin{document}
\newcommand{\nvminus}{NV$^{-}$}

\title{An \textit{ab-initio} effective solid state photoluminescence by frequency constraint of cluster calculation}

\author {Akib Karim}
    \email[Author to whom correspondence should be addressed: ]{akib.karim@rmit.edu.au}
    \affiliation{Quantum Photonics Laboratory, School of Engineering, RMIT University, Melbourne, VIC 3000, Australia}
    \affiliation{Centre for Quantum Computation and Communication Technology, School of Engineering, RMIT University, Melbourne, Victoria 3000, Australia}
\author{Igor Lyskov} 
\author{Salvy P. Russo}
\affiliation{ARC Centre of Excellence in Exciton Science, School of Science, RMIT University, Melbourne, VIC
3001 Australia}
\affiliation{Chemical and Quantum Physics, School of Science, RMIT University, Melbourne VIC 3001, Australia}
\author{Alberto Peruzzo}
\affiliation{Quantum Photonics Laboratory, School of Engineering, RMIT University, Melbourne, VIC 3000, Australia}
\affiliation{Centre for Quantum Computation and Communication Technology, School of Engineering, RMIT University, Melbourne, Victoria 3000, Australia}

\date{\today}

\begin{abstract}
Measuring the photoluminescence of defects in crystals is a common experimental technique for analysis and identification. However, current theoretical simulations typically require the simulation of a large number of atoms to eliminate finite size effects, which discourages computationally expensive excited state methods. We show how to extract the room-temperature photoluminescence spectra of defect centres in bulk from an \textit{ab-initio} simulation of a defect in small clusters. The finite size effect of small clusters manifests as strong coupling to low frequency vibrational modes. We find that removing vibrations below a cutoff frequency determined by constrained optimization returns the main features of the solid state photoluminescence spectrum. This strategy is illustrated for an NV$^{-}$ defect in diamond, presenting a connection between defects in solid state and clusters; the first vibrationally resolved \textit{ab-initio} photoluminescence spectrum of an NV$^{-}$ defect in a nanodiamond; and an alternative technique for simulating photoluminescence for solid state defects utilizing more accurate excited state methods.
\end{abstract}

\maketitle

\section{Introduction}

The creation and control of coherent single photons is an important step towards building next generation quantum technology. Photons represent an optimal controllable quantum system since they can encode quantum information that interacts minimally with the environment~\cite{Kok2007}. and can be easily transported and manipulated with free space and integrated photonic circuits~\cite{Bogdanov2017}. Traditional single photon sources based on spontaneous parametric downconversion are probabilistic in nature and therefore must be heralded or post-selected, severely limiting the scalability of photonic quantum technology~\cite{Eisaman2011}. Solid state systems, in particular defect centres in nanomaterials, are a promising architecture for deterministic single photon sources~\cite{Aharonovich2016}. These sources however, are generally limited by decoherence due to large coupling to vibrational states, which remains one of the biggest challenges in generating and collecting high quality photon emission from defect centres in nanomaterials~\cite{Albrecht2014}. The existence and frequency distribution of photon emission is observed experimentally in the photoluminescence (PL) spectrum, which is, therefore, used to investigate existing and discover new potential emitters. \textit{Ab-initio} calculation of PL can help to predict and explain electronic and vibrational properties of emitters. However, simulating PL for nanoscale systems is computationally hard due to the sheer number of parameters required to describe such systems, which limits the accuracy of results and the size of the systems that can be studied.

Traditionally, only electronic calculations were used for identifying and characterising emitters~\cite{Tran2016,Lu2012,Will1967}. Recently, the first \textit{ab-initio} calculation of the vibrational lineshape was calculated for a solid state emitter using the supercell method~\cite{Alkauskas2012, Alkauskas2014}, opening up the ability to identify emitters with the entire vibrationally resolved PL~\cite{Tawfik2017}. To get excited state properties, the method used $\Delta$-SCF, which is only mathematically proven for excited states with different symmetry than the ground state, or with the use of an orbital-specific exchange-correlation functional~\cite{Gorling1999}. While this applies to their test case of the NV$^{-}$ centre in diamond, this does not hold for all emitters, like the neutral nickel substitution in diamond Ni$_s^0$ ~\cite{Larico2009}.
It is preferable to treat excited states with Time Dependent Density Functional Theory (TD-DFT) as it gives formally exact excited state properties under similar approximations to DFT~\cite{Petersilka1996}. TD-DFT has been previously used to determine the electronic spectrum of NV$^{-}$ centres in nanodiamond~\cite{Gali2009b,gali2011b} and the absorption spectrum for SiV~\cite{Vlasov2014}.

Here we propose to calculate a PL spectrum for nanocrystals using the accuracy of electronic properties under TD-DFT with vibrational properties under DFT. To recover the PL spectrum for solid state crystals, the supercell method cannot be used because it is not compatible with nanocrystals in vacuum. We introduce a low frequency cutoff, which removes the vibrational modes which are not present in the solid-state.  
We validate the applicability of our method by calculating the PL spectrum of the NV$^{-}$ centre, since it is the most studied defect in the literature. 
Our method can be applied to general defect systems, and it is particularly attractive for systems which cannot be solved with previous methods, such as the systems with the same symmetry in the ground and excited states.\cite{Gorling1999}

\section{Methodology}
\begin{figure}
    \centering
    \includegraphics{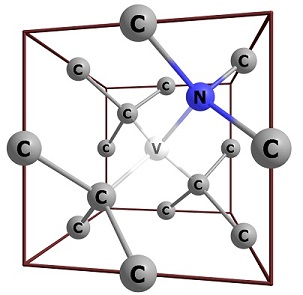}
    \caption{The structure of the NV$^{-}$ centre. The defect consists of a missing carbon, shown as V, with substituted nitrogen and extra electron. The nearest-neighbour atoms to the defect are shown. The C$_{197}$NH$_{140}$ carbon nanodiamond has three nearest neighbours to the defect.}
    \label{fig:geom}
\end{figure}

The NV$^{-}$ centre consists of a substituted nitrogen and a removed carbon from the diamond lattice. It exhibits C$_{3v}$ symmetry in the ground state and C$_{s}$ symmetry in the excited state. The defect is shown in Figure~\ref{fig:geom}. The electronic and vibrational structure of the NV$^{-}$ centre has been well studied. 

Electronically, the NV$^{-}$ centre consists of a triplet ground and excited state that are a linear combination of Slater determinants and cannot be simulated by DFT. Fortunately, the $m_{s}=1$ projection can be written as a Slater determinant~\cite{Gali2008} and has been shown to be sufficient for calculating transition energies~\cite{Gali2009b}. Configuration Interaction (CI) has been used to analyze these states~\cite{Bockstedte2018}. As CI is very computationally expensive, the TD-DFT level is a sufficient improvement on $\Delta$-SCF for this work as we only investigate the triplet to triplet transition.

Vibronically, the dynamical Jahn-Tellar effect of the defect has been characterized and accurate vibrational wavefunctions determined assuming a single effective frequency~\cite{Thiering2017}. The single effective frequency assumption is implicit in the Huang-Rhys model~\cite{Markham1959}. It has been shown~\cite{Thiering2017} that the \nvminus lineshape can be calculated under Franck-Condon approximations by relaxing symmetry from $C_{3v}$ to $C_{s}$. We follow that procedure for our study, so it is sufficient to ignore anharmonic effects and remain in the Franck-Condon picture.

\begin{figure}
    \centering
    \includegraphics{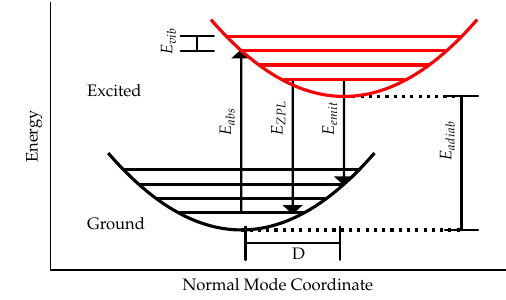}
    \caption{The displaced harmonic oscillator model (DHO) along one normal coordinate. The potential energy surface are shown as parabolae. Solid horizontal lines denote vibrational energy levels. The displacement on excitation along this mode is $D$. In the DHO model, the parabolae are identical but shifted by $D$. We calculate the ground state energy landscape with DFT to get normal modes. TD-DFT is only required to get a single point - the excited state energy minimum, which gives us the displacement vector}
    \label{fig:energ}
\end{figure}

To calculate the emission spectra from this model, we assume a displaced harmonic oscillator (DHO) model under the Franck-Condon approximation. The DHO model along one normal mode is shown in Figure~\ref{fig:energ}. $E_{vib}$ is the eigenfrequency of the mode and $D$ is the displacement of atoms on excitation along one normal co-ordinate. The vertical absorption and vertical emission ($E_{abs}, E_{emit}$) are the energy difference for transitions with no geometry change and are determined from TD-DFT on optimized ground and excited geometries respectively. $E_{ZPL}$ is known as the zero-phonon-line (ZPL) in solid state or $0\rightarrow0$ transition in finite systems. $E_{adiab}$ is the adiabatic energy, which is the difference between the energy of the relaxed ground and relaxed excited state geometries. 

In the DHO model, $E_{ZPL}$ is equal to $E_{adiab}$ as the ground and excited states have equal zero-point energies. $D$ for a vibrational mode defines the Partial Huang-Rhys (PHR) factor, which measures contribution of that mode to the PL. For details see supplementary material section S1: Analytic Derivation. The power of the displaced harmonic oscillator model is that the only parameters required to determine the PL are $E_{vib}$, $E_{adiab}$, and the PHR factors. 

Based on this, we modify the time-dependent approach to emission rate implemented by Etinski et al.~\cite{Etinski2014}. We assume dipole emission with a rate given by Fermi's Golden Rule. This rate contains an integral of a generating function over time. The Fourier transform of this function gives the PL. Our model is similar to the Huang-Rhys method used in Alkauskas et al.~\cite{Alkauskas2014} without implicitly assuming a single effective frequency. 

This calculates the PL for a cluster calculation under the DHO model. The cluster calculation is affected by finite-size effects which introduce coupling to normal modes of the surface atoms. This coupling does not exist in solid state because the crystal matrix will suppress movement along these degrees of freedom. Specifically for the solid state of $NV^{-}$, Webber et al.~\cite{Salvy2012}, using the supercell method with DFT, showed that the total vibrational density of states is dramatically suppressed at low frequencies.

In order to obtain the solid state PL spectrum from the nanocrystal calculation, we need to remove the contribution of modes at low frequencies. These can be identified from a constrained optimization that mimics solid state conditions by fixing the position of the outermost atoms. From this a low frequency cutoff can be identified by comparing unconstrained with constrained PHR factors. Removing the modes below the cutoff leads to a modified nanodiamond spectrum that mimics the solid state behaviour.

In summary, our method is as follows: first a nanocrystal with defect is constructed and DFT is performed to get the ground state optimized geometries. Then vibrational calculations under the harmonic approximation in DFT yield the $E_{vib}$ and normal coordinates as seen in Figure  \ref{fig:energ}. The ground state is also used with TD-DFT to yield the optimized excited geometry which gives $E_{adiab}$, $D$, and the PHR spectrum. A constrained TD-DFT geometry optimization then gives a constrained PHR spectrum, which allows us to identify the lower region of suppressed modes as well as the region of unaffected or amplified modes. The boundary between regions are the first frequencies where the constrained PHR value is less than or equal to the unconstrained. A cutoff value in this region is applied to the unconstrained PHR spectrum. The PHR spectrum with the ZPL from TD-DFT is input into the DHO model to get the PL.

This method requires one ground state geometry optimization DFT calculation; one normal mode calculation; two excited state geometry optimization TD-DFT calculations with and without constraints; and the final PL calculation. These are all standard procedures in computational chemistry. The exact equations we implemented for the PL calculation are given in the supplementary material, and can be implemented with minor modifications to the existing software, VIBES~\cite{Etinski2014}. 

Specifically, we construct a nanodiamond with composition C$_{197}$NH$_{140}$. This corresponds to third nearest neighbours to the defect. Previous electronic structure studies~\cite{Gali2011, Aittala2010} found that larger basis sets than double-$\zeta$ polarization do not increase accuracy significantly for geometry optimization. For all atoms, we use def2-SV(P)~\cite{Schafer1992}.

Gali et al.~\cite{Gali2011} found PBE0 was the optimal exchange-correlation function for electronic calculations of NV$^{-}$ under TD-DFT. As such we use PBE0 for all calculations.

Turbomole~\cite{Ahlrichs1989} was used to relax the geometry using DFT~\cite{Bauernschmitt1996} with C$_{3v}$ symmetry constraints. All excited properties were also calculated with TD-DFT~\cite{Grimme2002,Furche2005} in Turbomole using the same basis set and functional. The excited state optimization was performed under different constraints to analyze finite size effects. First, we performed unconstrained optimization under C$_{s}$. Then, we performed optimization with all CH and CH$_{2}$ groups constrained in ground state co-ordinates.

The program SNF~\cite{Neugebauer2002} was used to calculate the ground vibrational modes using finite difference with $0.01$ \AA~displacements. This consists of two displacements along three degrees of freedom for $338$ atoms, which requires $2028$ data points. Redundancy due to symmetry reduces this to $427$ single shot DFT calculations.

VIBES~\cite{Etinski2014} was used to calculate D and subsequently the PHR factors for the transition between the ground state and both constrained and unconstrained excited states. The difference of these was used to calculate boundaries between suppressed and unaffected regions. The cutoff frequency was found as the arithmetic mean between these frequencies.

We calculate vibrational lineshape using VIBES~\cite{Etinski2014} modified to implement the DHO model with a cutoff. As our model includes temperature effects (see supplementary materials section S1: Analytic Derivation for details), the temperature was set to $300$ K. The correlation function was evaluated over $300$ fs at $16384$ intervals. To replicate experimental line broadening, the spectrum was convoluted with a Gaussian of width $200$ cm$^{-1}$, which is approximately the experimental FWHM of the ZPL in Figure~\ref{fig:pl}.

\section{Results}
\begin{figure}
\flushleft{(a)}
\includegraphics[clip]{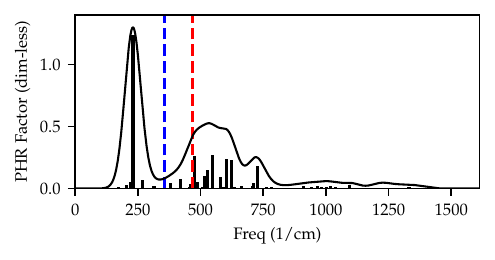}

\vspace{-0.1cm}
\flushleft{(b)}
\includegraphics[clip]{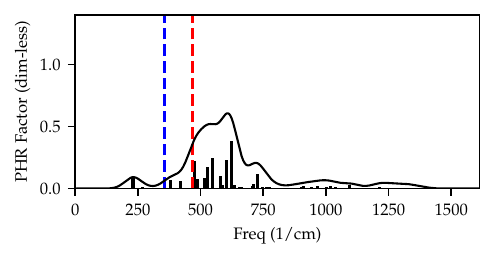}

\vspace{-0.1cm}
\flushleft{(c)}
\includegraphics[clip]{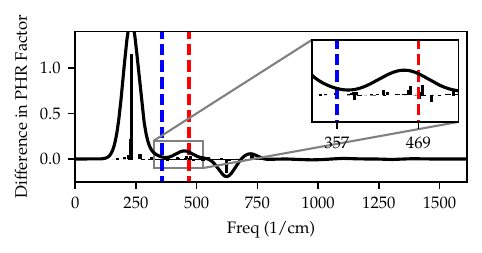}
\caption{Partial Huang-Rhys (PHR) factors as functions of frequency for (a) unconstrained, (b) constrained excited states and (c) difference between (a) and (b). PHR factors indicate the contribution of a mode to the PL. PHR factors are given as delta functions. The solid line shows PHR factors with Gaussian broadening to demonstrate distinct regions. The difference plot in (c) shows which regions are suppressed (positive) or unaffected (zero or negative). The blue and red vertical dotted lines show frequencies where the constrained PHR value in (b) are first less than or equal to the unconstrained value in (a) after the large positive region and before the large negative region respectively. The cutoffs chosen in this region can recover peaks of the PL spectrum.}
\label{fig:vib}
\end{figure}

The adiabatic energy, $E_{adiab}$ for the C$_{197}$NH$_{140}$ nanodiamond under TD-DFT is $1.945$ eV. This is consistent with the experimental ZPL value of $1.945$ eV~\cite{Bradac2012}. While this appears extraordinarily accurate, it will depend on the shape of the simulated nanodiamond (see supplementary material section S2: Other Sized Nanodiamonds for details). Despite this, the results are an improvement on $\Delta$-SCF, which reported $1.757$ eV under PBE and $2.035$ eV under HSE. These are consistent with previous TD-DFT results, which found $E_{emit}$ of $2.2$ eV compared to our $2.1$ eV~\cite{Gali2011}.

Figure~\ref{fig:vib}(a) shows the PHR factors as a function of the frequency of their respective normal mode for a C$_{197}$NH$_{140}$ nanodiamond. The PHR factors show the amount of coupling of a normal mode to the emission. The vibrational modes show three distinct regions. At low frequency, most modes involve large displacements from all atoms; at $480$-$650$ $cm^{-1}$, the vibrational modes have the largest amplitude with the closest atoms to the defect, as shown in Figure~\ref{fig:geom}; higher frequency modes involve smaller subregions, for instance modes around $1500$ $cm^{-1}$ are exclusively C-H bending modes but cannot be seen in Figure \ref{fig:energ}, since they do not couple to emission and have a PHR factor of zero. The middle region at $532$ $cm^{-1}$ corresponds to the $65$ meV peak identified by Alkauskas et al.~\cite{Alkauskas2014}.

In Figure~\ref{fig:vib}(b) we report the PHR factors obtained by constraining the CH and CH$_{2}$ groups, and by plotting the difference between constrained and unconstrained factors in Figure~\ref{fig:vib}(c), we can see that at low frequencies (on the left of the blue boundary) the presence of large positive peaks indicate that the coupling to vibrational modes are present in the nanodiamond but not in bulk, whereas at high frequencies (on the right of the red boundary) the vibrational modes equally couple to the transition for both bulk and nanodiamond. The region within the boundaries therefore separates the unique properties of the nanodiamond from those in common with bulk. Eliminating the vibrational modes with frequencies below this cutoff region should return the bulk behaviour.
In Figure~\ref{fig:HR-other}, we compare the unconstrained and constrained PHR factors for three different nanocrystal shapes, and show that the suppression of the low frequency peaks occurs under constraining across all shapes.

We can also compare the PHR factors with vibrational data obtained for bulk $NV^{-}$ as reported in Webber et al.~\cite{Salvy2012}. As the authors report in Figure 2, the total density of states begins to increase at $200$  $cm^{-1}$, illustrating a lack of vibrational states at low frequencies. Comparison with Figure~\ref{fig:vib}(a), shows that the nanodiamond has large coupling to frequencies in this same region. The coupling of low frequency modes we see in the nanodiamond can, thus, be considered unique to the nanodiamond and are not present in bulk. This motivates our use of a low frequency cutoff to eliminate these in order to imitate the solid-state behaviour. Note that we derive the value of our cutoff from nanodiamond calculations with constraints, not from solid state data. 
Similarly, consider the bulk vibrational states projected onto nitrogen also reported in~\cite{Salvy2012}. These indicate vibrational modes that involve nitrogen and are likely to be important for defect transitions. These modes begin around $350$ $cm^{-1}$ and peak around $600$ $cm^{-1}$. The peak corresponds to a similar cluster of peaks in the PHR factors around $532$ $cm^{-1}$ in Figure~\ref{fig:vib}(a). Furthermore, the lowest frequency nitrogen modes are similar to our lower bound for the cutoff at $357$ $cm^{-1}$ calculated from comparing unconstrained to constrained PHR factors. The constraint of outer carbons would eliminate vibrational modes that involve the edge of the nanodiamond, but not modes of the defect, in particular nitrogen. It is unsurprising, then, that we see correspondence between our cutoff and the $350$ $cm^{-1}$ edge of nitrogen projected modes in bulk.
The two distinct peaks of Figure~\ref{fig:vib}(a) demonstrate why the assumption of a single effective frequency from the Huang-Rhys model is invalid. In comparison, PL for solid state NV$^{-}$ has been reproduced under the mean frequency assumption~\cite{Alkauskas2014}. In that sense, our low frequency cutoff reproduces solid state spectra from cluster calculations by mimicking a single effective frequency around $532$ $cm^{-1}$.

\begin{figure*}
    \centering
    \includegraphics[width=1.0\linewidth]{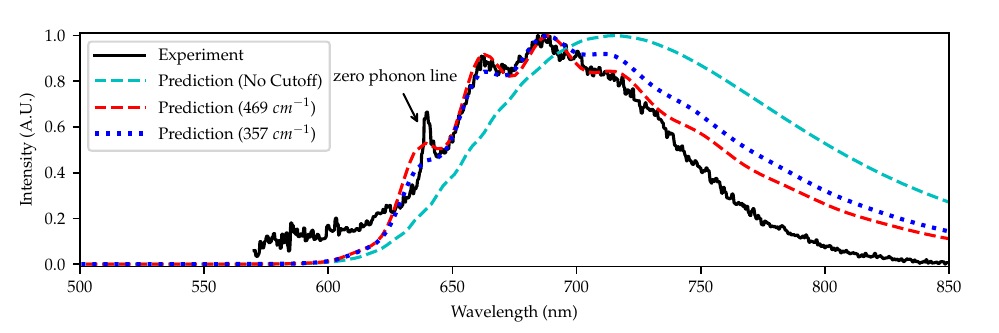}
    \caption{Photoluminescence spectrum of an NV$^{-}$ centre in a C$_{197}$NH$_{140}$ nanodiamond with frequency cutoffs at $357$ $cm^{-1}$ (blue dotted) and $469$ $cm^{-1}$ (red dashed) and without any cutoff applied (cyan). Experimental spectrum of solid state NV$^{-}$ is reproduced with permission from Aslam et al.~\cite{Aslam2013}}
    \label{fig:pl}
\end{figure*}

The predicted PL spectra are given in Figure~\ref{fig:pl} (Figure S2(b), extends the results for larger range of cutoff frequencies). The spectra with no cutoff is the \textit{ab-initio} spectra for a C$_{197}$NH$_{140}$ nanodiamond in vacuum based on our displaced harmonic oscillator model. Experimental spectra for small size nanodiamonds is not available because \nvminus centers in such small crystals are unlikely to form~\cite{Bolshedvorskii2017,Bradac2009}. In comparison, both PL spectra with cutoff match the peaks of the solid state experimental spectrum and recreate the lineshape of the sideband well for both $357$ and $469$ $cm^{-1}$ cutoffs. The spectrum without cutoff has larger intensity at long wavelengths than the experiment. The spectrum based on cutoff at $357$ $cm^{-1}$ has a longer tail than the one at $469$ $cm^{-1}$, which shows that the tail is due to contribution from the large low energy peak in figure~\ref{fig:vib}(a) as well as from the smaller peak around $400$ $cm^{-1}$. The cutoff spectra is less intense at long wavelengths, but is still higher than experiment. 

We are not able to recreate the long wavelength tail of the spectrum. This is likely due to the inelegance of using a strict cutoff on low frequency modes. It is likely that higher frequency modes must be modified. This idea is consistent with Figure~\ref{fig:vib} where the vibrational mode around $645$ $cm^{-1}$ increases in the constrained case. Furthermore, the constrained calculation is not an exact simulation of the solid state. We freeze the CH and CH$_{2}$ groups when in reality they are only suppressed by a diamond lattice. These can be explored in future work. Regardless, we have established that clusters introduce a region of low frequency vibrations that couple to excitation and cause the PL to have much higher intensity at long wavelengths compared to solid state PL. 

We have demonstrated an ab-initio method for calculating the photoluminescence spectrum of solid state emitters from cluster calculations. We have shown that finite size effects from cluster calculations appear as part of a low frequency peak in the Huang-Rhys spectrum. We have shown that a low frequency cutoff recovers the solid state PL spectrum. Furthermore, we have demonstrated the first vibrationally resolved PL spectrum for a nanodiamond as well as the first spectrum utilising TD-DFT for excited state properties, which has fewer restrictions than $\Delta$-SCF. This method was demonstrated with NV$^{-}$ and found to match experimental PL peaks but loses intensity from the ZPL to long wavelengths.
A more general approach to predict the value of the cutoff will require future work involving comparisons across multiple defects.

Our method enables the calculation of the photoluminescence spectrum for defects of arbitrary symmetry between the ground and excited states, beyond the NV$^{-}$ centre. This constitutes a valuable tool for understanding and predicting solid state single photon emitter properties.

See supplementary material for analytic derivation of the PL spectrum; electronic and vibrational results for C$_{121}$NH$_{100}$ and  C$_{145}$NH$_{100}$; and spectra generated by varying the cutoff frequency as well as from constrained data without cutoff.

The data that support the findings of this study are available from the corresponding author upon reasonable request.

\begin{acknowledgments}
A.P. acknowledges a RMIT University Vice-Chancellor’s Senior Research Fellowship; ARC DECRA Fellowship (No: DE140101700), and Google Faculty Research Award. This work was supported by the Australian Government through the Australian Research Council under the Centre of Excellence scheme (No: CE170100012, CE170100026). It was also supported by computational resources provided by the Australian Government through the National Computational Infrastructure Facility and the Pawsey Supercomputer Centre.
\end{acknowledgments}

\setcounter{table}{0}
\renewcommand{\thetable}{S\arabic{table}}%
\setcounter{figure}{0}
\renewcommand{\thefigure}{S\arabic{figure}}%
\renewcommand{\theequation}{S\arabic{equation}}
\setcounter{section}{0}
\renewcommand{\thesection}{S\arabic{section}}

\section{S1: Analytic Derivation}\label{Theory}
\subsection{Derivation of Photoluminescence Spectrum}
To calculate vibronic intensities, we use the displaced harmonic oscillator model. This model assumes separate electronic and vibrational degrees of freedom with parabolic potential energy surfaces for vibration. The potential energy surfaces for vibration in the excited electronic state are identical to the ground state but displaced with regards to atomic position. This is shown in Figure~\ref{fig:energ} in the main text. Henceforth, we follow the derivations in Etinski et al. and Markham~\cite{Markham1959,Etinski2014}

The Hamiltonian thus consists of a standard Born-Oppenheimer Hamiltonian for electronic states and a vibrational portion dependent on the nuclear parameters $q_{i}$ given as two displaced harmonic oscillators along each normal mode coordinate $i$:
\begin{equation}
    H = \sum_{i}^{N} (H_{g,i} + H_{e,i}),
\end{equation}

where,

\begin{equation}
    H_{g,i} = \frac{p^2}{2m} + \frac{1}{2}m\omega_{i}^{2}q_{i}^{2}
\end{equation}
and
\begin{equation}
    H_{e,i} = \frac{p^2}{2m} + \frac{1}{2}m\omega_{i}^{2}(q_{i}-d_{i})^{2},
\end{equation}

where p is momentum, m is mass, $\omega_{i}$ is the frequency of the harmonic oscillator along the $i$th normal coordinate, $d_{i}$ is the displacement on excitation for the $i$th normal coordinate.

Assuming dipole emission, we then use Fermi's golden rule to write the probability of emission from an excited state $\ket{E_{ex}, v_{ex}}$ to $\ket{E_{gr}, v_{gr}}$ as:

\begin{equation}
    T = \frac{2\pi}{\hbar}\bra{E_{ex}, v_{ex}} \Hat{\mu} \ket{E_{gr}, v_{gr}},
\end{equation}

where we now use the Franck-Condon approximation to assume the dipole operator has no dependence on nuclear co-ordinates, which allows us to pull the vibrational states out:

\begin{equation}
    T = \frac{2\pi}{\hbar}\mu_{E,G} |\bra{v_{ex}}\ket{v_{gr}}|^2 \delta(E_e - E_g),
\end{equation}
where $\mu_{E,G}$ is the electronic transition dipole moment.

We can now rewrite the delta function in integral form and rewrite the vibrational overlap as the integral over Hermite polynomials to give:

\begin{equation}
    T = \frac{1}{\hbar}\mu_{E,G}  \int^{\infty}_{-\infty} G(t) dt,
\end{equation}

with generating function $G(t)$:
\begin{widetext}
\begin{equation}
    G(t) = e^{-\Delta{E}it}\Pi_{j} 2 sinh(\frac{\beta\omega_{j}}{2}) \Sigma_{n} e^{-(n_j + \frac{1}{2})(i\omega_{j}t)} \chi_{j}(q_{j}) \chi_{j}(\Bar{q_{j}}) 
    \Sigma_{n'} e^{-(n'_j + \frac{1}{2})(\beta\omega_{j} - i\omega_{j}t)} \chi_{j}(q_{ex,j}) \chi_{j}(\Bar{q_{ex,j}}),
\end{equation}
\end{widetext}
where $\chi_{j}(q_{j})$ denotes the Hermite polynomial solution along the $j$th normal mode coordinate direction and $q_{ex,j} = q_{j} + d_{j}$.

Under the displaced harmonic oscillator, the purely electronic transition, $\Delta E$, is identical to the adiabatic energy since both electronic states have the same zero point energy.

The generating function can be grossly simplified using Mehler's formula for Hermite polynomials $\chi_j$:

\begin{equation}
\begin{split}
\Sigma_{n} e^{-\Sigma_j (n_j + \frac{1}{2})\xi_{j}} \chi_{n_1}(q_1)  ... \chi_{n_N}(q_N) \chi_{n_1}(\Bar{q_1}) ... \chi_{n_N}(\Bar{q_N})\\
= (2\pi)^{-\frac{N}{2}} \sqrt{\det(S^{-1} \Omega)}
\exp \Big(-\frac{1}{4} (Q+\Bar{Q})^\dag \Omega B(Q+ \Bar{Q})\\
-\frac{1}{4}(Q -\Bar{Q})^\dag \Omega B^{-1} (Q-\Bar{Q}) \Big).
\end{split}
\end{equation}

$\Omega$, $S$ and $B$ are diagonal matrices with elements $\omega$, $\sinh(\xi_{j})$ and $\tanh(\xi_{j}/2)$ respectively and the vector $Q = (q_{1},q_{2},...,q_{N})$ and $\Bar{Q}$ denote some displacement of the molecule along its normal mode coordinates. For the ground state, $\xi_{j} = i\omega_{j} t$; for the excited state, $\xi_{j} = \omega_{j}(\beta-it)$ where $\beta$ is the Boltzmann factor defined by: $\frac{\hbar}{kT}$. This substituted into $G(t)$ for ground and excited harmonic oscillators gives:

\begin{equation}\label{eq:QQ}
\begin{split}
    G(t) = (2\pi)^{-N} \sqrt{\det(2S_{\beta}^2 S^{-1}_{gr}S^{-1}_{ex} \Omega_{gr} \Omega_{ex})} 
    \\\int exp(-\frac{1}{4} ((Q_{gr}+ \Bar{Q_{gr}})^\dag \Omega_{gr} B_{gr}(Q_{gr}+ \Bar{Q_{gr}})+
    \\(Q_{gr} -\Bar{Q_{gr}})^\dag \Omega_{gr} B_{gr}^{-1} (Q_{gr}-\Bar{Q_{gr}}) +
    \\ (Q_{ex}+ \Bar{Q_{ex}})^\dag \Omega_{ex} B_{ex}(Q_{ex}+\Bar{Q_{ex}})+
    \\ (Q_{ex} -\Bar{Q_{ex}})^\dag \Omega_{ex} B_{ex}^{-1} (Q_{ex}-\Bar{Q_{ex}})) d^N Q d^N \Bar{Q},
\end{split}
\end{equation}

where $S_\beta$ is the diagonal matrix with entries $\sinh(\frac{\beta\omega_i}{2})$

From equations $2$ and $3$, we can see that the normal and excited normal modes differ only in translation by the vector $D = (d_1, d_2,..., d_N)$:

\begin{equation}
    Q_{ex} = Q_{gr} + D.
\end{equation}

Furthermore, the diagonal matrix $\Omega$ will be the same for ground and excited state, so we will drop the subscript.

we substitute this into equation~\ref{eq:QQ} and transform into co-ordinates $X=Q+\Bar{Q}$ and $Y=Q-\Bar{Q}$ to obtain:

\begin{equation}
\begin{split}
    G(t) = (-4\pi)^{-N} \sqrt{det(2S_{\beta}^2 S^{-1}_{gr}S^{-1}_{ex} \Omega^2)} exp(-D^{\dag} \Omega B_{ex} D) \\
    \int \exp(-\frac{1}{2} (X^\dag \Omega (B_{gr} + B_{ex}) X) - D^\dag \Omega B_{ex} X ) dX^N\\
    \int \exp(-\frac{1}{2} (Y^\dag \Omega (B_{ex}^{-1} + B_{gr}^{-1}) Y)) dY^N.
\end{split}
\end{equation}

These correspond to two vector Gaussian integrals which can be evaluated analytically to give:

\begin{equation}
\begin{split}
    G(t) = \sqrt{\frac{\det(2S_{\beta}^2 S^{-1}_{gr}S^{-1}_{ex} \Omega^2)}{det(\Omega(B_{gr} + B_{ex}))\det(\Omega(B_{gr}^{-1} + B_{ex}^{-1}))}} \\
    e^{-D^\dag \Omega B_{ex} D} e^{\frac{1}{2}D^\dag \Omega B_{ex} (\Omega B_{gr} + \Omega B_{ex})^{-1} B_{ex}^\dag \Omega^\dag D},
    \end{split}
\end{equation}

We now use some basic hyperbolic trigonometric identities and the fact the matrices $\Omega$ and $B$ are diagonal to simplify as:

\begin{equation}\label{eq:vibes}
    G(t) = \exp(-\frac{1}{2}D^\dag \Omega \frac{B_{ex}^2 + 2B_{ex}B_{gr}}{B_{gr} + B_{ex}} D).
\end{equation}

The spectrum is then given by the Fourier transform of the generating function. This formula is implemented in a modified version of Vibes~\cite{Etinski2014}

\subsection{Effect of Cutoff frequency}

To see the effect of the cutoff frequency, we write the terms in the exponential in index notation:
\begin{multline}\label{eq:cutoff}
    -\sum_{j=1}^{N}\bigg( \\\frac{\tanh^2(\omega_{j}(\beta-it)/2) + 2\tanh(\omega_{j}(\beta-it)/2)\tanh(i\omega_{j} t/2)}{\tanh(\omega_{j}(\beta-it)/2) + \tanh(i\omega_{j} t/2)}\\ \frac{d_{j}^2 \omega_{j}}{2}\bigg).
\end{multline}

The generating function is now explicitly a sum over frequency modes. In order to remove the effect of a particular mode, we remove its corresponding term from the sum. Let us reorder the sum such that $\omega_{j} > \omega_{j+1}$ for all $j$. We see that a low frequency cutoff of $k$ modes is simply equation~\ref{eq:cutoff} with $N$ replaced with $N-k$.

The low frequency cutoff can be considered equivalent to constraining the degrees of freedom related to the low frequency region. Visual inspection of the associated modes reveals this to be largely vibrations that involve the entire nanodiamond.

\subsection{Huang-Rhys Constants}
The analysis above is exact for the spectrum of dipole emission from the displaced harmonic oscillator Hamiltonian under the Franck-Condon approximation. 

To get the Huang-Rhys constants, we make the additional assumption that the hyperbolic functions all oscillate at an identical mean frequency, $\omega$. This allows us to factor out the hyperbolic terms from the sum over modes in equation~\ref{eq:cutoff}:

\begin{multline}
    -\frac{\tanh^2(\omega(\beta-it)/2) + 2\tanh(\omega(\beta-it)/2)\tanh(i\omega t/2)}{\tanh(\omega(\beta-it)/2) + \tanh(i\omega t/2)} \\\Sigma_{i} \frac{d_{i}^2 \omega_{i}}{2}.
\end{multline}

The correlation function is only a function of each individual eigenfrequency in this sum term. We now introduce the famous Huang-Rhys constant as:

\begin{equation}
    S = \Sigma_{i} \frac{d_{i}^2 \omega_{i}}{2}.
\end{equation}

\begin{figure*}

\subfloat[][C$_{121}$NH$_{100}$ unconstrained]{\includegraphics{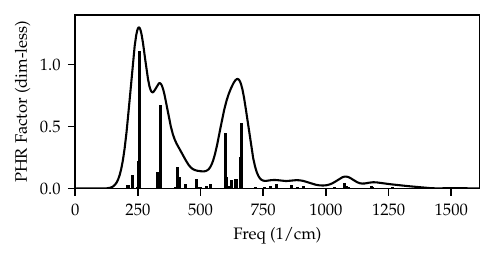}}
\subfloat[][C$_{121}$NH$_{100}$ constrained]{\includegraphics{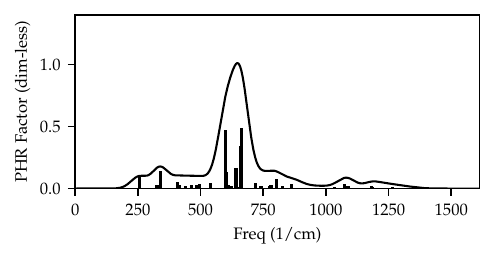}}

\subfloat[][C$_{145}$NH$_{100}$ unconstrained]{\includegraphics{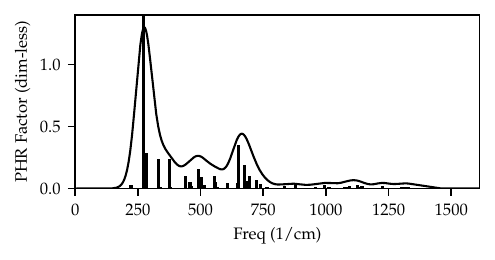}}
\subfloat[][C$_{145}$NH$_{100}$ constrained]{\includegraphics{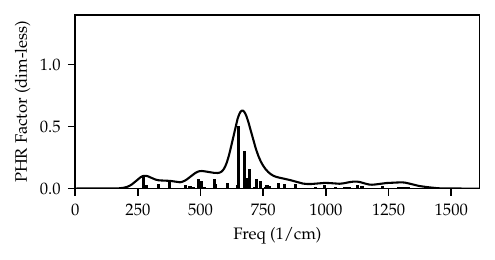}}

\subfloat[][C$_{197}$NH$_{140}$ unconstrained]{\includegraphics{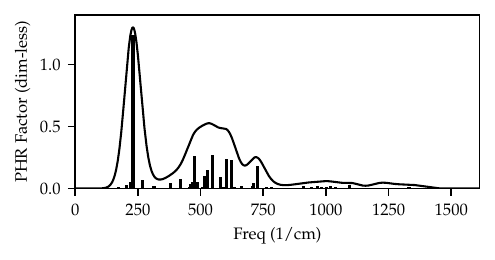}}
\subfloat[][C$_{197}$NH$_{140}$ constrained]{\includegraphics{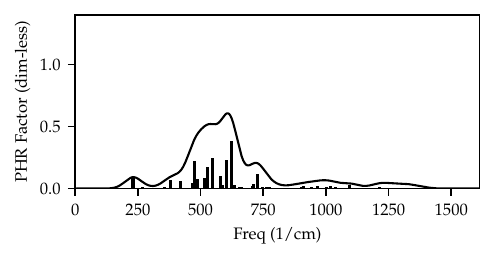}}

\caption{Partial Huang-Rhys (HR) factors as functions of frequency for constrained and unconstrained C$_{121}$NH$_{100}$, C$_{145}$NH$_{100}$, and C$_{197}$NH$_{140}$ diamond. The cluster of normal modes just after $400$~$cm^{-1}$ correspond to the additional modes included in the cutoff from Figure~\ref{fig:vib} in the main text. Note that Figure~\ref{fig:HR-other} is aligned with unconstrained simulations on the left and constrained simulations on the right, which is the transposition from Figure~\ref{fig:vib}.}
\label{fig:HR-other}
\end{figure*}

In models such as the Fröhlich hamiltonian or the CC model this quantity has physical interpretation as the number of vibrational modes passed during a transition. In other words, 

\begin{equation}
    S = (E_{emit} - E_{ZPL} - E_{ZPE}) / E_{vib},
\end{equation}

where $E_{emit}$ is the emission energy, $E_{ZPL}$ is the zero phonon line energy, $E_{vib}$ is $\hbar\omega$, and $E_{ZPE}$ is the zero point energy.

This is not the case in our model. To get temperature effects, we assume a Boltzmann distribution in excited state vibrational levels. S only has physical interpretation if we emit only from the lowest vibrational state of the excited state. Furthermore, we do not use the crucial mean frequency assumption.

Rather, it is useful to consider the partial Huang-Rhys constants, i.e.

\begin{equation}\label{eq:partials}
    S_{i} = \frac{d_{i}^2 \omega_{i}}{2},
\end{equation}

\begin{figure*}

\flushleft{(a)}
\vspace{1.0cm}
\includegraphics[clip]{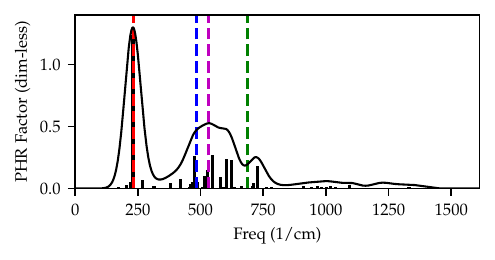} \centering

\vspace{-1.1cm}
(b)
\includegraphics[clip]{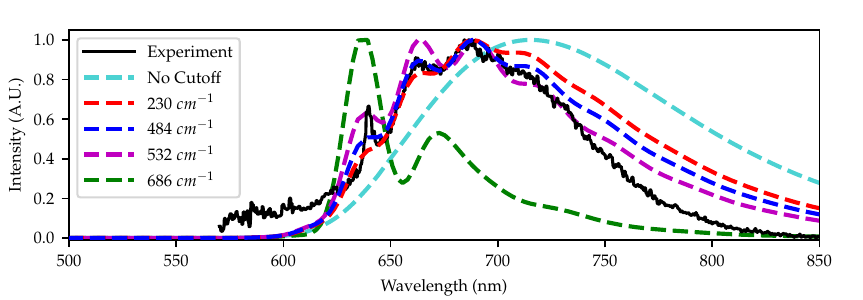}

\caption{(a) Partial Huang-Rhys (PHR) factors as functions of frequency with multiple cutoffs shown as vertical lines. (b) Corresponding PL spectra for each of these cutoffs. The color of the vertical lines in (a) correspond to the color of the PL in (b).}

\label{fig:cutoff}
\end{figure*}

which are the only terms in the correlation function that depend on $d$. These can be used as a metric for contribution of each normal mode to the excitation spectra. Furthermore, they can be used as comparison to other models in which the total Huang-Rhys constant has physical interpretation. The partial Huang-Rhys constants are plotted in Figure~\ref{fig:vib}. 

\section{S2: Other sized nanodiamonds}\label{Electronics}

\begin{table}[b]
\caption{Electronic energies (as depicted in Figure~\ref{fig:energ}) and total Huang-Rhys constants for the three shapes of nanodiamond tested.}
\begin{tabular}{|l|l|l|l|l}

\hline
Composition & C$_{121}$NH$_{100}$ & C$_{145}$NH$_{100}$ & C$_{197}$NH$_{140}$  \\ \hline
Diameter (nm)  & $1.03$       & $1.09$       & $1.31$\\ \hline
E$_{abs}$ (eV)   & $2.17$       & $2.27$       & $2.10$       \\
E$_{emit}$ (eV)  & $1.60$       & $1.64$       & $1.56$             \\
E$_{adiab}$ (eV) & $1.99$       & $2.17$       & $1.95$       \\ \hline
S (dim-less)   & $5.05$       & $4.80$       & $4.32$
   \\ \hline
\end{tabular}
\end{table}

The method was performed on three nanodiamond sizes or shapes: C$_{121}$NH$_{100}$, C$_{145}$NH$_{100}$ and C$_{197}$NH$_{140}$. In this section, firstly, we outline a table of the energies calculated from TD-DFT and total Huang-Rhys constants. Secondly, we show the partial Huang-Rhys factors as a function of frequency for the three nanodiamonds. Figure~\ref{fig:energ} in the main text illustrates each energy with reference to the displaced harmonic oscillator model.

As C$_{197}$NH$_{140}$ recovers the zero phonon line, we chose to use it for the photoluminescence calculation in the main text. 

The partial Huang-Rhys factors for these nanodiamonds are shown in Figure~\ref{fig:HR-other} in the supplementary material. They demonstrate the inadequacy of a mean frequency model to explain vibronic coupling. Furthermore, Figure~\ref{fig:HR-other}(c), the C$_{145}$N$H_{100}$ nanodiamond unconstrained spectra, illustrates how the cluster of peaks around $400$~$cm^{-1}$ are distinct from the $532$~$cm^{-1}$ peak. This is a result of the shape difference in the C$_{145}$N$H_{100}$ nanodiamond.

The integral of the PHR spectra is called the Huang-Rhys constant. The Huang-Rhys constant does not have obvious physical significance in our DHO model, but we can compare the value to literature. We calculate a value of $4.32$ for the spectra with the cutoff. Previous simulations give values of $3.67$ and experiments give $3.8$~\cite{Alkauskas2014}. In the Huang-Rhys model, this value can be interpreted as how many effective vibrational levels are passed in a vertical transition. There is no physical interpretation to relate the HR factor to our PL spectrum.

\section{S3: Varying the cutoff}

\begin{figure*}
    \centering
    \includegraphics{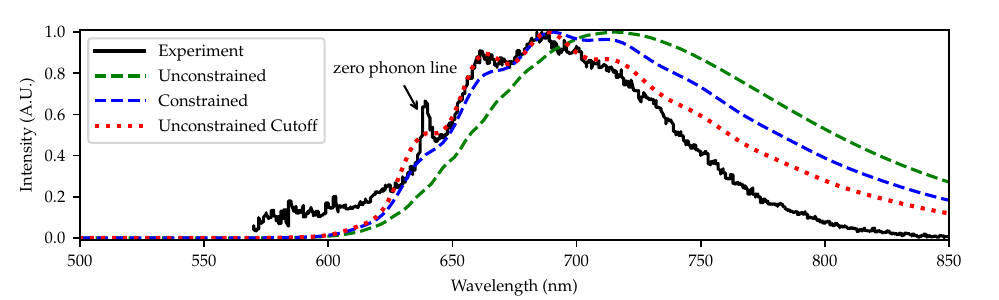}
    \caption{The photoluminescence spectrum of NV$^{-}$ from the unconstrained calculation with and without a cutoff; and the constrained data without a cutoff. Experimental data from Aslam et al.~\cite{Aslam2013} }
    \label{fig:PL_con}
\end{figure*}

The cutoff frequency is calculated ab-initio by comparing the PHR spectra for constrained and unconstrained excited states. However, it is possible to generate PL spectra for different cutoffs to identify the contribution to the PL spectra by different regions of the PHR spectrum.

From Figure~\ref{fig:cutoff}, we can see that using the \textit{ab-initio} cutoff attains the characteristic peaks of the solid-state PL spectrum. Without the cutoff, the low energy/long wavelength contributions in the PHR spectrum dominate over these peaks and make the peaks hard to resolve. As the cutoff is increased, we start to resolve the characteristic peaks. Past $484$~$cm^{-1}$, we can see that we begin to lose emission from the characteristic peaks in the longer wavelengths, in particular the peak at $690$~nm. The emission is instead in the ZPL, increasing its intensity. It is interesting to note that the $532$~$cm^{-1}$ cutoff recreates the relative ZPL intensity, but incorrectly predicts the relative heights of the remaining peaks.

The \textit{ab-initio} cutoff therefore removes low energy (long wavelength) modes present in the cluster calculation but not in the solid state while preserving the vibrational modes required to get the characteristic peaks in the solid state.

\section{S4: PL of constrained spectra without cutoff}

We use a constrained optimization calculation to identify which normal modes would be suppressed in a crystalline environment. As discussed in the main text, this is not a realistic simulation of a crystal. Calculating the PL spectrum from this raw constrained calculation can help identify why a low frequency cutoff is necessary.

We calculated the PL spectrum from the constrained Partial Huang-Rhys factors of Figure~\ref{fig:vib}b). This can be seen in supplementary material Figure~\ref{fig:PL_con}. The constrained optimisation does not entirely eliminate the low frequency peak at $230$~$cm^{-1}$ (See Figure 3a in the paper). It has a Partial Huang-Rhys factor of $0.1$, which is not negligible, especially as the highest coupled mode at $621$~$cm^{-1}$ has a PHR factor of $0.37$. This coupling to the low frequency region is enough to add significant emission to the tail of the spectra at wavelengths longer than $720$~nm. It also has reduced emission from the zero phonon line and first vibrational peak around $670$~nm. The low frequency cutoff is necessary to completely remove coupling from low frequency modes.

\bibliography{NVPL}

\begin{thebibliography}{35}%
\makeatletter
\providecommand \@ifxundefined [1]{%
 \@ifx{#1\undefined}
}%
\providecommand \@ifnum [1]{%
 \ifnum #1\expandafter \@firstoftwo
 \else \expandafter \@secondoftwo
 \fi
}%
\providecommand \@ifx [1]{%
 \ifx #1\expandafter \@firstoftwo
 \else \expandafter \@secondoftwo
 \fi
}%
\providecommand \natexlab [1]{#1}%
\providecommand \enquote  [1]{``#1''}%
\providecommand \bibnamefont  [1]{#1}%
\providecommand \bibfnamefont [1]{#1}%
\providecommand \citenamefont [1]{#1}%
\providecommand \href@noop [0]{\@secondoftwo}%
\providecommand \href [0]{\begingroup \@sanitize@url \@href}%
\providecommand \@href[1]{\@@startlink{#1}\@@href}%
\providecommand \@@href[1]{\endgroup#1\@@endlink}%
\providecommand \@sanitize@url [0]{\catcode `\\12\catcode `\$12\catcode
  `\&12\catcode `\#12\catcode `\^12\catcode `\_12\catcode `\%12\relax}%
\providecommand \@@startlink[1]{}%
\providecommand \@@endlink[0]{}%
\providecommand \url  [0]{\begingroup\@sanitize@url \@url }%
\providecommand \@url [1]{\endgroup\@href {#1}{\urlprefix }}%
\providecommand \urlprefix  [0]{URL }%
\providecommand \Eprint [0]{\href }%
\providecommand \doibase [0]{http://dx.doi.org/}%
\providecommand \selectlanguage [0]{\@gobble}%
\providecommand \bibinfo  [0]{\@secondoftwo}%
\providecommand \bibfield  [0]{\@secondoftwo}%
\providecommand \translation [1]{[#1]}%
\providecommand \BibitemOpen [0]{}%
\providecommand \bibitemStop [0]{}%
\providecommand \bibitemNoStop [0]{.\EOS\space}%
\providecommand \EOS [0]{\spacefactor3000\relax}%
\providecommand \BibitemShut  [1]{\csname bibitem#1\endcsname}%
\let\auto@bib@innerbib\@empty
\bibitem [{\citenamefont {Kok}\ \emph {et~al.}(2007)\citenamefont {Kok},
  \citenamefont {Munro}, \citenamefont {Nemoto}, \citenamefont {Ralph},
  \citenamefont {Dowling},\ and\ \citenamefont {Milburn}}]{Kok2007}%
  \BibitemOpen
  \bibfield  {author} {\bibinfo {author} {\bibfnamefont {P.}~\bibnamefont
  {Kok}}, \bibinfo {author} {\bibfnamefont {W.~J.}\ \bibnamefont {Munro}},
  \bibinfo {author} {\bibfnamefont {K.}~\bibnamefont {Nemoto}}, \bibinfo
  {author} {\bibfnamefont {T.~C.}\ \bibnamefont {Ralph}}, \bibinfo {author}
  {\bibfnamefont {J.~P.}\ \bibnamefont {Dowling}}, \ and\ \bibinfo {author}
  {\bibfnamefont {G.~J.}\ \bibnamefont {Milburn}},\ }\href {\doibase
  10.1103/RevModPhys.79.135} {\bibfield  {journal} {\bibinfo  {journal} {Rev.
  Mod. Phys.}\ }\textbf {\bibinfo {volume} {79}},\ \bibinfo {pages} {135}
  (\bibinfo {year} {2007})}\BibitemShut {NoStop}%
\bibitem [{\citenamefont {Bogdanov}\ \emph {et~al.}(2017)\citenamefont
  {Bogdanov}, \citenamefont {Shalaginov}, \citenamefont {Boltasseva},\ and\
  \citenamefont {Shalaev}}]{Bogdanov2017}%
  \BibitemOpen
  \bibfield  {author} {\bibinfo {author} {\bibfnamefont {S.}~\bibnamefont
  {Bogdanov}}, \bibinfo {author} {\bibfnamefont {M.~Y.}\ \bibnamefont
  {Shalaginov}}, \bibinfo {author} {\bibfnamefont {A.}~\bibnamefont
  {Boltasseva}}, \ and\ \bibinfo {author} {\bibfnamefont {V.~M.}\ \bibnamefont
  {Shalaev}},\ }\href {\doibase 10.1364/ome.7.000111} {\bibfield  {journal}
  {\bibinfo  {journal} {Opt. Mater. Express}\ }\textbf {\bibinfo {volume}
  {7}},\ \bibinfo {pages} {111} (\bibinfo {year} {2017})}\BibitemShut {NoStop}%
\bibitem [{\citenamefont {Eisaman}\ \emph {et~al.}(2011)\citenamefont
  {Eisaman}, \citenamefont {Fan}, \citenamefont {Migdall},\ and\ \citenamefont
  {Polyakov}}]{Eisaman2011}%
  \BibitemOpen
  \bibfield  {author} {\bibinfo {author} {\bibfnamefont {M.~D.}\ \bibnamefont
  {Eisaman}}, \bibinfo {author} {\bibfnamefont {J.}~\bibnamefont {Fan}},
  \bibinfo {author} {\bibfnamefont {A.}~\bibnamefont {Migdall}}, \ and\
  \bibinfo {author} {\bibfnamefont {S.~V.}\ \bibnamefont {Polyakov}},\
  }\href@noop {} {\bibfield  {journal} {\bibinfo  {journal} {Rev. Sci.
  Instrum.}\ }\textbf {\bibinfo {volume} {82}} (\bibinfo {year}
  {2011})}\BibitemShut {NoStop}%
\bibitem [{\citenamefont {Aharonovich}\ \emph {et~al.}(2016)\citenamefont
  {Aharonovich}, \citenamefont {Englund},\ and\ \citenamefont
  {Toth}}]{Aharonovich2016}%
  \BibitemOpen
  \bibfield  {author} {\bibinfo {author} {\bibfnamefont {I.}~\bibnamefont
  {Aharonovich}}, \bibinfo {author} {\bibfnamefont {D.}~\bibnamefont
  {Englund}}, \ and\ \bibinfo {author} {\bibfnamefont {M.}~\bibnamefont
  {Toth}},\ }\href {\doibase 10.1038/nphoton.2016.186} {\bibfield  {journal}
  {\bibinfo  {journal} {Nat. Photonics}\ }\textbf {\bibinfo {volume} {10}},\
  \bibinfo {pages} {631} (\bibinfo {year} {2016})}\BibitemShut {NoStop}%
\bibitem [{\citenamefont {Albrecht}\ \emph {et~al.}(2014)\citenamefont
  {Albrecht}, \citenamefont {Bommer}, \citenamefont {Pauly}, \citenamefont
  {M{\"{u}}cklich}, \citenamefont {Schell}, \citenamefont {Engel},
  \citenamefont {Schr{\"{o}}der}, \citenamefont {Benson}, \citenamefont
  {Reichel},\ and\ \citenamefont {Becher}}]{Albrecht2014}%
  \BibitemOpen
  \bibfield  {author} {\bibinfo {author} {\bibfnamefont {R.}~\bibnamefont
  {Albrecht}}, \bibinfo {author} {\bibfnamefont {A.}~\bibnamefont {Bommer}},
  \bibinfo {author} {\bibfnamefont {C.}~\bibnamefont {Pauly}}, \bibinfo
  {author} {\bibfnamefont {F.}~\bibnamefont {M{\"{u}}cklich}}, \bibinfo
  {author} {\bibfnamefont {A.~W.}\ \bibnamefont {Schell}}, \bibinfo {author}
  {\bibfnamefont {P.}~\bibnamefont {Engel}}, \bibinfo {author} {\bibfnamefont
  {T.}~\bibnamefont {Schr{\"{o}}der}}, \bibinfo {author} {\bibfnamefont
  {O.}~\bibnamefont {Benson}}, \bibinfo {author} {\bibfnamefont
  {J.}~\bibnamefont {Reichel}}, \ and\ \bibinfo {author} {\bibfnamefont
  {C.}~\bibnamefont {Becher}},\ }\href {\doibase 10.1063/1.4893612} {\bibfield
  {journal} {\bibinfo  {journal} {Appl. Phys. Lett.}\ }\textbf {\bibinfo
  {volume} {105}},\ \bibinfo {pages} {073113} (\bibinfo {year}
  {2014})}\BibitemShut {NoStop}%
\bibitem [{\citenamefont {Tran}\ \emph {et~al.}(2016)\citenamefont {Tran},
  \citenamefont {Bray}, \citenamefont {Ford}, \citenamefont {Toth},\ and\
  \citenamefont {Aharonovich}}]{Tran2016}%
  \BibitemOpen
  \bibfield  {author} {\bibinfo {author} {\bibfnamefont {T.~T.}\ \bibnamefont
  {Tran}}, \bibinfo {author} {\bibfnamefont {K.}~\bibnamefont {Bray}}, \bibinfo
  {author} {\bibfnamefont {M.~J.}\ \bibnamefont {Ford}}, \bibinfo {author}
  {\bibfnamefont {M.}~\bibnamefont {Toth}}, \ and\ \bibinfo {author}
  {\bibfnamefont {I.}~\bibnamefont {Aharonovich}},\ }\href {\doibase
  10.1038/nnano.2015.242} {\bibfield  {journal} {\bibinfo  {journal} {Nat.
  Nanotechnol.}\ }\textbf {\bibinfo {volume} {11}},\ \bibinfo {pages} {37}
  (\bibinfo {year} {2016})}\BibitemShut {NoStop}%
\bibitem [{\citenamefont {Lu}\ \emph {et~al.}(2012)\citenamefont {Lu},
  \citenamefont {Lin}, \citenamefont {Chou}, \citenamefont {Peng},
  \citenamefont {Lo},\ and\ \citenamefont {Cheng}}]{Lu2012}%
  \BibitemOpen
  \bibfield  {author} {\bibinfo {author} {\bibfnamefont {H.~C.}\ \bibnamefont
  {Lu}}, \bibinfo {author} {\bibfnamefont {M.~Y.}\ \bibnamefont {Lin}},
  \bibinfo {author} {\bibfnamefont {S.~L.}\ \bibnamefont {Chou}}, \bibinfo
  {author} {\bibfnamefont {Y.~C.}\ \bibnamefont {Peng}}, \bibinfo {author}
  {\bibfnamefont {J.~I.}\ \bibnamefont {Lo}}, \ and\ \bibinfo {author}
  {\bibfnamefont {B.~M.}\ \bibnamefont {Cheng}},\ }\href@noop {} {\bibfield
  {journal} {\bibinfo  {journal} {Anal Chem}\ }\textbf {\bibinfo {volume}
  {84}},\ \bibinfo {pages} {9596} (\bibinfo {year} {2012})}\BibitemShut
  {NoStop}%
\bibitem [{\citenamefont {Williams}\ and\ \citenamefont
  {Blacknall}(1967)}]{Will1967}%
  \BibitemOpen
  \bibfield  {author} {\bibinfo {author} {\bibfnamefont {E.}~\bibnamefont
  {Williams}}\ and\ \bibinfo {author} {\bibfnamefont {D.}~\bibnamefont
  {Blacknall}},\ }\href@noop {} {\bibfield  {journal} {\bibinfo  {journal}
  {Trans. Met. Sue. AIME}\ }\textbf {\bibinfo {volume} {239}},\ \bibinfo
  {pages} {387} (\bibinfo {year} {1967})}\BibitemShut {NoStop}%
\bibitem [{\citenamefont {Alkauskas}\ \emph {et~al.}(2012)\citenamefont
  {Alkauskas}, \citenamefont {Lyons}, \citenamefont {Steiauf},\ and\
  \citenamefont {{Van De Walle}}}]{Alkauskas2012}%
  \BibitemOpen
  \bibfield  {author} {\bibinfo {author} {\bibfnamefont {A.}~\bibnamefont
  {Alkauskas}}, \bibinfo {author} {\bibfnamefont {J.~L.}\ \bibnamefont
  {Lyons}}, \bibinfo {author} {\bibfnamefont {D.}~\bibnamefont {Steiauf}}, \
  and\ \bibinfo {author} {\bibfnamefont {C.~G.}\ \bibnamefont {{Van De
  Walle}}},\ }\href {\doibase 10.1103/PhysRevLett.109.267401} {\bibfield
  {journal} {\bibinfo  {journal} {Phys. Rev. Lett.}\ }\textbf {\bibinfo
  {volume} {109}},\ \bibinfo {pages} {1} (\bibinfo {year} {2012})}\BibitemShut
  {NoStop}%
\bibitem [{\citenamefont {Alkauskas}\ \emph {et~al.}(2014)\citenamefont
  {Alkauskas}, \citenamefont {Buckley}, \citenamefont {Awschalom},\ and\
  \citenamefont {{Van De Walle}}}]{Alkauskas2014}%
  \BibitemOpen
  \bibfield  {author} {\bibinfo {author} {\bibfnamefont {A.}~\bibnamefont
  {Alkauskas}}, \bibinfo {author} {\bibfnamefont {B.~B.}\ \bibnamefont
  {Buckley}}, \bibinfo {author} {\bibfnamefont {D.~D.}\ \bibnamefont
  {Awschalom}}, \ and\ \bibinfo {author} {\bibfnamefont {C.~G.}\ \bibnamefont
  {{Van De Walle}}},\ }\href@noop {} {\bibfield  {journal} {\bibinfo  {journal}
  {New J. Phys.}\ }\textbf {\bibinfo {volume} {16}},\ \bibinfo {pages} {073026}
  (\bibinfo {year} {2014})}\BibitemShut {NoStop}%
\bibitem [{\citenamefont {Pustovarov}\ \emph {et~al.}(2011)\citenamefont
  {Pustovarov}, \citenamefont {Perevalov}, \citenamefont {Gritsenko},
  \citenamefont {Smirnova},\ and\ \citenamefont {Yelisseyev}}]{Tawfik2017}%
  \BibitemOpen
  \bibfield  {author} {\bibinfo {author} {\bibfnamefont {V.~A.}\ \bibnamefont
  {Pustovarov}}, \bibinfo {author} {\bibfnamefont {T.~V.}\ \bibnamefont
  {Perevalov}}, \bibinfo {author} {\bibfnamefont {V.~A.}\ \bibnamefont
  {Gritsenko}}, \bibinfo {author} {\bibfnamefont {T.~P.}\ \bibnamefont
  {Smirnova}}, \ and\ \bibinfo {author} {\bibfnamefont {A.~P.}\ \bibnamefont
  {Yelisseyev}},\ }\href {\doibase 10.1016/j.tsf.2011.04.014} {\bibfield
  {journal} {\bibinfo  {journal} {Thin Solid Films}\ }\textbf {\bibinfo
  {volume} {519}},\ \bibinfo {pages} {6319} (\bibinfo {year}
  {2011})}\BibitemShut {NoStop}%
\bibitem [{\citenamefont {G{\"{o}}rling}(1999)}]{Gorling1999}%
  \BibitemOpen
  \bibfield  {author} {\bibinfo {author} {\bibfnamefont {A.}~\bibnamefont
  {G{\"{o}}rling}},\ }\href {\doibase 10.1103/PhysRevA.59.3359} {\bibfield
  {journal} {\bibinfo  {journal} {Phys. Rev. A}\ }\textbf {\bibinfo {volume}
  {59}},\ \bibinfo {pages} {3359} (\bibinfo {year} {1999})}\BibitemShut
  {NoStop}%
\bibitem [{\citenamefont {Larico}\ \emph {et~al.}(2009)\citenamefont {Larico},
  \citenamefont {Justo}, \citenamefont {Machado},\ and\ \citenamefont
  {Assali}}]{Larico2009}%
  \BibitemOpen
  \bibfield  {author} {\bibinfo {author} {\bibfnamefont {R.}~\bibnamefont
  {Larico}}, \bibinfo {author} {\bibfnamefont {J.~F.}\ \bibnamefont {Justo}},
  \bibinfo {author} {\bibfnamefont {W.~V.}\ \bibnamefont {Machado}}, \ and\
  \bibinfo {author} {\bibfnamefont {L.~V.}\ \bibnamefont {Assali}},\
  }\href@noop {} {\bibfield  {journal} {\bibinfo  {journal} {Phys. Rev. B}\
  }\textbf {\bibinfo {volume} {79}} (\bibinfo {year} {2009})}\BibitemShut
  {NoStop}%
\bibitem [{\citenamefont {Petersilka}\ \emph {et~al.}(1996)\citenamefont
  {Petersilka}, \citenamefont {Gossmann},\ and\ \citenamefont
  {Gross}}]{Petersilka1996}%
  \BibitemOpen
  \bibfield  {author} {\bibinfo {author} {\bibfnamefont {M.}~\bibnamefont
  {Petersilka}}, \bibinfo {author} {\bibfnamefont {U.~J.}\ \bibnamefont
  {Gossmann}}, \ and\ \bibinfo {author} {\bibfnamefont {E.~K.}\ \bibnamefont
  {Gross}},\ }\href {\doibase 10.1103/PhysRevLett.76.1212} {\bibfield
  {journal} {\bibinfo  {journal} {Phys. Rev. Lett.}\ }\textbf {\bibinfo
  {volume} {76}},\ \bibinfo {pages} {1212} (\bibinfo {year}
  {1996})}\BibitemShut {NoStop}%
\bibitem [{\citenamefont {Gali}\ \emph {et~al.}(2009)\citenamefont {Gali},
  \citenamefont {Janz\'en}, \citenamefont {De\'ak}, \citenamefont {Kresse},\
  and\ \citenamefont {Kaxiras}}]{Gali2009b}%
  \BibitemOpen
  \bibfield  {author} {\bibinfo {author} {\bibfnamefont {A.}~\bibnamefont
  {Gali}}, \bibinfo {author} {\bibfnamefont {E.}~\bibnamefont {Janz\'en}},
  \bibinfo {author} {\bibfnamefont {P.}~\bibnamefont {De\'ak}}, \bibinfo
  {author} {\bibfnamefont {G.}~\bibnamefont {Kresse}}, \ and\ \bibinfo {author}
  {\bibfnamefont {E.}~\bibnamefont {Kaxiras}},\ }\href {\doibase
  10.1103/PhysRevLett.103.186404} {\bibfield  {journal} {\bibinfo  {journal}
  {Phys. Rev. Lett.}\ }\textbf {\bibinfo {volume} {103}},\ \bibinfo {pages}
  {186404} (\bibinfo {year} {2009})}\BibitemShut {NoStop}%
\bibitem [{\citenamefont {Thiering}\ and\ \citenamefont
  {Gali}(2011)}]{gali2011b}%
  \BibitemOpen
  \bibfield  {author} {\bibinfo {author} {\bibfnamefont {G.}~\bibnamefont
  {Thiering}}\ and\ \bibinfo {author} {\bibfnamefont {A.}~\bibnamefont
  {Gali}},\ }\href {\doibase 10.1002/pssb.201046254} {\bibfield  {journal}
  {\bibinfo  {journal} {physica status solidi (b)}\ }\textbf {\bibinfo {volume}
  {248}},\ \bibinfo {pages} {1337 } (\bibinfo {year} {2011})}\BibitemShut
  {NoStop}%
\bibitem [{\citenamefont {Vlasov}\ \emph {et~al.}(2014)\citenamefont {Vlasov},
  \citenamefont {Shiryaev}, \citenamefont {Rendler}, \citenamefont {Steinert},
  \citenamefont {Lee}, \citenamefont {Antonov}, \citenamefont
  {V{\"{o}}r{\"{o}}s}, \citenamefont {Jelezko}, \citenamefont {Fisenko},
  \citenamefont {Semjonova}, \citenamefont {Biskupek}, \citenamefont {Kaiser},
  \citenamefont {Lebedev}, \citenamefont {Sildos}, \citenamefont {Hemmer},
  \citenamefont {Konov}, \citenamefont {Gali},\ and\ \citenamefont
  {Wrachtrup}}]{Vlasov2014}%
  \BibitemOpen
  \bibfield  {author} {\bibinfo {author} {\bibfnamefont {I.~I.}\ \bibnamefont
  {Vlasov}}, \bibinfo {author} {\bibfnamefont {A.~A.}\ \bibnamefont
  {Shiryaev}}, \bibinfo {author} {\bibfnamefont {T.}~\bibnamefont {Rendler}},
  \bibinfo {author} {\bibfnamefont {S.}~\bibnamefont {Steinert}}, \bibinfo
  {author} {\bibfnamefont {S.~Y.}\ \bibnamefont {Lee}}, \bibinfo {author}
  {\bibfnamefont {D.}~\bibnamefont {Antonov}}, \bibinfo {author} {\bibfnamefont
  {M.}~\bibnamefont {V{\"{o}}r{\"{o}}s}}, \bibinfo {author} {\bibfnamefont
  {F.}~\bibnamefont {Jelezko}}, \bibinfo {author} {\bibfnamefont {A.~V.}\
  \bibnamefont {Fisenko}}, \bibinfo {author} {\bibfnamefont {L.~F.}\
  \bibnamefont {Semjonova}}, \bibinfo {author} {\bibfnamefont {J.}~\bibnamefont
  {Biskupek}}, \bibinfo {author} {\bibfnamefont {U.}~\bibnamefont {Kaiser}},
  \bibinfo {author} {\bibfnamefont {O.~I.}\ \bibnamefont {Lebedev}}, \bibinfo
  {author} {\bibfnamefont {I.}~\bibnamefont {Sildos}}, \bibinfo {author}
  {\bibfnamefont {P.~R.}\ \bibnamefont {Hemmer}}, \bibinfo {author}
  {\bibfnamefont {V.~I.}\ \bibnamefont {Konov}}, \bibinfo {author}
  {\bibfnamefont {A.}~\bibnamefont {Gali}}, \ and\ \bibinfo {author}
  {\bibfnamefont {J.}~\bibnamefont {Wrachtrup}},\ }\href {\doibase
  10.1038/nnano.2013.255} {\bibfield  {journal} {\bibinfo  {journal} {Nat.
  Nanotechnol.}\ }\textbf {\bibinfo {volume} {9}},\ \bibinfo {pages} {54}
  (\bibinfo {year} {2014})}\BibitemShut {NoStop}%
\bibitem [{\citenamefont {Gali}\ \emph {et~al.}(2008)\citenamefont {Gali},
  \citenamefont {Fyta},\ and\ \citenamefont {Kaxiras}}]{Gali2008}%
  \BibitemOpen
  \bibfield  {author} {\bibinfo {author} {\bibfnamefont {A.}~\bibnamefont
  {Gali}}, \bibinfo {author} {\bibfnamefont {M.}~\bibnamefont {Fyta}}, \ and\
  \bibinfo {author} {\bibfnamefont {E.}~\bibnamefont {Kaxiras}},\ }\href
  {\doibase 10.1103/PhysRevB.77.155206} {\bibfield  {journal} {\bibinfo
  {journal} {Phys. Rev. B}\ }\textbf {\bibinfo {volume} {77}},\ \bibinfo
  {pages} {1} (\bibinfo {year} {2008})}\BibitemShut {NoStop}%
\bibitem [{\citenamefont {Bockstedte}\ \emph {et~al.}(2018)\citenamefont
  {Bockstedte}, \citenamefont {Sch{\"{u}}tz}, \citenamefont {Garratt},
  \citenamefont {Iv{\'{a}}dy},\ and\ \citenamefont {Gali}}]{Bockstedte2018}%
  \BibitemOpen
  \bibfield  {author} {\bibinfo {author} {\bibfnamefont {M.}~\bibnamefont
  {Bockstedte}}, \bibinfo {author} {\bibfnamefont {F.}~\bibnamefont
  {Sch{\"{u}}tz}}, \bibinfo {author} {\bibfnamefont {T.}~\bibnamefont
  {Garratt}}, \bibinfo {author} {\bibfnamefont {V.}~\bibnamefont
  {Iv{\'{a}}dy}}, \ and\ \bibinfo {author} {\bibfnamefont {A.}~\bibnamefont
  {Gali}},\ }\href@noop {} {\bibfield  {journal} {\bibinfo  {journal} {npj
  Quantum Mater.}\ }\textbf {\bibinfo {volume} {3}},\ \bibinfo {pages} {1}
  (\bibinfo {year} {2018})}\BibitemShut {NoStop}%
\bibitem [{\citenamefont {Thiering}\ and\ \citenamefont
  {Gali}(2017)}]{Thiering2017}%
  \BibitemOpen
  \bibfield  {author} {\bibinfo {author} {\bibfnamefont {G.}~\bibnamefont
  {Thiering}}\ and\ \bibinfo {author} {\bibfnamefont {A.}~\bibnamefont
  {Gali}},\ }\href {\doibase 10.1103/PhysRevB.96.081115} {\bibfield  {journal}
  {\bibinfo  {journal} {Phys. Rev. B}\ }\textbf {\bibinfo {volume} {96}},\
  \bibinfo {pages} {081115} (\bibinfo {year} {2017})}\BibitemShut {NoStop}%
\bibitem [{\citenamefont {Markham}(1959)}]{Markham1959}%
  \BibitemOpen
  \bibfield  {author} {\bibinfo {author} {\bibfnamefont {J.~J.}\ \bibnamefont
  {Markham}},\ }\href {\doibase 10.1103/RevModPhys.31.956} {\bibfield
  {journal} {\bibinfo  {journal} {Rev. Mod. Phys.}\ }\textbf {\bibinfo {volume}
  {31}},\ \bibinfo {pages} {956} (\bibinfo {year} {1959})}\BibitemShut
  {NoStop}%
\bibitem [{\citenamefont {Etinski}\ \emph {et~al.}(2014)\citenamefont
  {Etinski}, \citenamefont {Rai-Constapel},\ and\ \citenamefont
  {Marian}}]{Etinski2014}%
  \BibitemOpen
  \bibfield  {author} {\bibinfo {author} {\bibfnamefont {M.}~\bibnamefont
  {Etinski}}, \bibinfo {author} {\bibfnamefont {V.}~\bibnamefont
  {Rai-Constapel}}, \ and\ \bibinfo {author} {\bibfnamefont {C.~M.}\
  \bibnamefont {Marian}},\ }\href {\doibase 10.1063/1.4868484} {\bibfield
  {journal} {\bibinfo  {journal} {J. Chem. Phys.}\ }\textbf {\bibinfo {volume}
  {140}},\ \bibinfo {pages} {114104} (\bibinfo {year} {2014})}\BibitemShut
  {NoStop}%
\bibitem [{\citenamefont {Webber}\ \emph {et~al.}(2012)\citenamefont {Webber},
  \citenamefont {Per}, \citenamefont {Drumm}, \citenamefont {Hollenberg},\ and\
  \citenamefont {Russo}}]{Salvy2012}%
  \BibitemOpen
  \bibfield  {author} {\bibinfo {author} {\bibfnamefont {B.~T.}\ \bibnamefont
  {Webber}}, \bibinfo {author} {\bibfnamefont {M.~C.}\ \bibnamefont {Per}},
  \bibinfo {author} {\bibfnamefont {D.~W.}\ \bibnamefont {Drumm}}, \bibinfo
  {author} {\bibfnamefont {L.~C.~L.}\ \bibnamefont {Hollenberg}}, \ and\
  \bibinfo {author} {\bibfnamefont {S.~P.}\ \bibnamefont {Russo}},\ }\href
  {\doibase 10.1103/PhysRevB.85.014102} {\bibfield  {journal} {\bibinfo
  {journal} {Phys. Rev. B}\ }\textbf {\bibinfo {volume} {85}},\ \bibinfo
  {pages} {014102} (\bibinfo {year} {2012})}\BibitemShut {NoStop}%
\bibitem [{\citenamefont {Gali}\ \emph {et~al.}(2011)\citenamefont {Gali},
  \citenamefont {Simon},\ and\ \citenamefont {Lowther}}]{Gali2011}%
  \BibitemOpen
  \bibfield  {author} {\bibinfo {author} {\bibfnamefont {A.}~\bibnamefont
  {Gali}}, \bibinfo {author} {\bibfnamefont {T.}~\bibnamefont {Simon}}, \ and\
  \bibinfo {author} {\bibfnamefont {J.~E.}\ \bibnamefont {Lowther}},\
  }\href@noop {} {\bibfield  {journal} {\bibinfo  {journal} {New J. Phys.}\
  }\textbf {\bibinfo {volume} {13}},\ \bibinfo {pages} {025016} (\bibinfo
  {year} {2011})}\BibitemShut {NoStop}%
\bibitem [{\citenamefont {Aittala}\ \emph {et~al.}(2010)\citenamefont
  {Aittala}, \citenamefont {Cramariuc}, \citenamefont {Hukka}, \citenamefont
  {Vasilescu}, \citenamefont {Bandula},\ and\ \citenamefont
  {Lemmetyinen}}]{Aittala2010}%
  \BibitemOpen
  \bibfield  {author} {\bibinfo {author} {\bibfnamefont {P.~J.}\ \bibnamefont
  {Aittala}}, \bibinfo {author} {\bibfnamefont {O.}~\bibnamefont {Cramariuc}},
  \bibinfo {author} {\bibfnamefont {T.~I.}\ \bibnamefont {Hukka}}, \bibinfo
  {author} {\bibfnamefont {M.}~\bibnamefont {Vasilescu}}, \bibinfo {author}
  {\bibfnamefont {R.}~\bibnamefont {Bandula}}, \ and\ \bibinfo {author}
  {\bibfnamefont {H.}~\bibnamefont {Lemmetyinen}},\ }\href {\doibase
  10.1021/jp9104536} {\bibfield  {journal} {\bibinfo  {journal} {J. Phys. Chem.
  A}\ }\textbf {\bibinfo {volume} {114}},\ \bibinfo {pages} {7094} (\bibinfo
  {year} {2010})}\BibitemShut {NoStop}%
\bibitem [{\citenamefont {Sch{\"{a}}fer}\ \emph {et~al.}(1992)\citenamefont
  {Sch{\"{a}}fer}, \citenamefont {Horn},\ and\ \citenamefont
  {Ahlrichs}}]{Schafer1992}%
  \BibitemOpen
  \bibfield  {author} {\bibinfo {author} {\bibfnamefont {A.}~\bibnamefont
  {Sch{\"{a}}fer}}, \bibinfo {author} {\bibfnamefont {H.}~\bibnamefont {Horn}},
  \ and\ \bibinfo {author} {\bibfnamefont {R.}~\bibnamefont {Ahlrichs}},\
  }\href {\doibase 10.1063/1.463096} {\bibfield  {journal} {\bibinfo  {journal}
  {J. Chem. Phys.}\ }\textbf {\bibinfo {volume} {97}},\ \bibinfo {pages} {2571}
  (\bibinfo {year} {1992})}\BibitemShut {NoStop}%
\bibitem [{\citenamefont {Ahlrichs}\ \emph {et~al.}(1989)\citenamefont
  {Ahlrichs}, \citenamefont {B{\"{a}}r}, \citenamefont {H{\"{a}}ser},
  \citenamefont {Horn},\ and\ \citenamefont {K{\"{o}}lmel}}]{Ahlrichs1989}%
  \BibitemOpen
  \bibfield  {author} {\bibinfo {author} {\bibfnamefont {R.}~\bibnamefont
  {Ahlrichs}}, \bibinfo {author} {\bibfnamefont {M.}~\bibnamefont {B{\"{a}}r}},
  \bibinfo {author} {\bibfnamefont {M.}~\bibnamefont {H{\"{a}}ser}}, \bibinfo
  {author} {\bibfnamefont {H.}~\bibnamefont {Horn}}, \ and\ \bibinfo {author}
  {\bibfnamefont {C.}~\bibnamefont {K{\"{o}}lmel}},\ }\href {\doibase
  10.1016/0009-2614(89)85118-8} {\bibfield  {journal} {\bibinfo  {journal}
  {Chem. Phys. Lett.}\ }\textbf {\bibinfo {volume} {162}},\ \bibinfo {pages}
  {165} (\bibinfo {year} {1989})}\BibitemShut {NoStop}%
\bibitem [{\citenamefont {Bauernschmitt}\ and\ \citenamefont
  {Ahlrichs}(1996)}]{Bauernschmitt1996}%
  \BibitemOpen
  \bibfield  {author} {\bibinfo {author} {\bibfnamefont {R.}~\bibnamefont
  {Bauernschmitt}}\ and\ \bibinfo {author} {\bibfnamefont {R.}~\bibnamefont
  {Ahlrichs}},\ }\href {\doibase 10.1016/0009-2614(96)00440-X} {\bibfield
  {journal} {\bibinfo  {journal} {Chem. Phys. Lett.}\ }\textbf {\bibinfo
  {volume} {256}},\ \bibinfo {pages} {454} (\bibinfo {year}
  {1996})}\BibitemShut {NoStop}%
\bibitem [{\citenamefont {Grimme}\ \emph {et~al.}(2002)\citenamefont {Grimme},
  \citenamefont {Furche},\ and\ \citenamefont {Ahlrichs}}]{Grimme2002}%
  \BibitemOpen
  \bibfield  {author} {\bibinfo {author} {\bibfnamefont {S.}~\bibnamefont
  {Grimme}}, \bibinfo {author} {\bibfnamefont {F.}~\bibnamefont {Furche}}, \
  and\ \bibinfo {author} {\bibfnamefont {R.}~\bibnamefont {Ahlrichs}},\ }\href
  {\doibase 10.1016/S0009-2614(02)00975-2} {\bibfield  {journal} {\bibinfo
  {journal} {Chem. Phys. Lett.}\ }\textbf {\bibinfo {volume} {361}},\ \bibinfo
  {pages} {321} (\bibinfo {year} {2002})}\BibitemShut {NoStop}%
\bibitem [{\citenamefont {Furche}\ and\ \citenamefont
  {Rappoport}(2005)}]{Furche2005}%
  \BibitemOpen
  \bibfield  {author} {\bibinfo {author} {\bibfnamefont {F.}~\bibnamefont
  {Furche}}\ and\ \bibinfo {author} {\bibfnamefont {D.}~\bibnamefont
  {Rappoport}},\ }\href {\doibase 10.1016/S1380-7323(05)80020-2} {\bibfield
  {journal} {\bibinfo  {journal} {Theor. Comput. Chem.}\ }\textbf {\bibinfo
  {volume} {16}},\ \bibinfo {pages} {93} (\bibinfo {year} {2005})}\BibitemShut
  {NoStop}%
\bibitem [{\citenamefont {Neugebauer}\ \emph {et~al.}(2002)\citenamefont
  {Neugebauer}, \citenamefont {Reiher}, \citenamefont {Kind},\ and\
  \citenamefont {Hess}}]{Neugebauer2002}%
  \BibitemOpen
  \bibfield  {author} {\bibinfo {author} {\bibfnamefont {J.}~\bibnamefont
  {Neugebauer}}, \bibinfo {author} {\bibfnamefont {M.}~\bibnamefont {Reiher}},
  \bibinfo {author} {\bibfnamefont {C.}~\bibnamefont {Kind}}, \ and\ \bibinfo
  {author} {\bibfnamefont {B.~A.}\ \bibnamefont {Hess}},\ }\href {\doibase
  10.1002/jcc.10089} {\bibfield  {journal} {\bibinfo  {journal} {J. Comput.
  Chem.}\ }\textbf {\bibinfo {volume} {23}},\ \bibinfo {pages} {895} (\bibinfo
  {year} {2002})}\BibitemShut {NoStop}%
\bibitem [{\citenamefont {Bradac}(2012)}]{Bradac2012}%
  \BibitemOpen
  \bibfield  {author} {\bibinfo {author} {\bibfnamefont {C.}~\bibnamefont
  {Bradac}},\ }\href@noop {} {Ph.D. thesis},\ \bibinfo  {school} {Macquarie
  University} (\bibinfo {year} {2012})\BibitemShut {NoStop}%
\bibitem [{\citenamefont {Aslam}\ \emph {et~al.}(2013)\citenamefont {Aslam},
  \citenamefont {Waldherr}, \citenamefont {Neumann}, \citenamefont {Jelezko},\
  and\ \citenamefont {Wrachtrup}}]{Aslam2013}%
  \BibitemOpen
  \bibfield  {author} {\bibinfo {author} {\bibfnamefont {N.}~\bibnamefont
  {Aslam}}, \bibinfo {author} {\bibfnamefont {G.}~\bibnamefont {Waldherr}},
  \bibinfo {author} {\bibfnamefont {P.}~\bibnamefont {Neumann}}, \bibinfo
  {author} {\bibfnamefont {F.}~\bibnamefont {Jelezko}}, \ and\ \bibinfo
  {author} {\bibfnamefont {J.}~\bibnamefont {Wrachtrup}},\ }\href@noop {}
  {\bibfield  {journal} {\bibinfo  {journal} {New J. Phys.}\ }\textbf {\bibinfo
  {volume} {15}},\ \bibinfo {pages} {013064} (\bibinfo {year}
  {2013})}\BibitemShut {NoStop}%
\bibitem [{\citenamefont {Bolshedvorskii}\ \emph {et~al.}(2017)\citenamefont
  {Bolshedvorskii}, \citenamefont {Vorobyov}, \citenamefont {Soshenko},
  \citenamefont {Shershulin}, \citenamefont {Javadzade}, \citenamefont
  {Zeleneev}, \citenamefont {Komrakova}, \citenamefont {Sorokin}, \citenamefont
  {Belobrov}, \citenamefont {Smolyaninov},\ and\ \citenamefont
  {Akimov}}]{Bolshedvorskii2017}%
  \BibitemOpen
  \bibfield  {author} {\bibinfo {author} {\bibfnamefont {S.~V.}\ \bibnamefont
  {Bolshedvorskii}}, \bibinfo {author} {\bibfnamefont {V.~V.}\ \bibnamefont
  {Vorobyov}}, \bibinfo {author} {\bibfnamefont {V.~V.}\ \bibnamefont
  {Soshenko}}, \bibinfo {author} {\bibfnamefont {V.~A.}\ \bibnamefont
  {Shershulin}}, \bibinfo {author} {\bibfnamefont {J.}~\bibnamefont
  {Javadzade}}, \bibinfo {author} {\bibfnamefont {A.~I.}\ \bibnamefont
  {Zeleneev}}, \bibinfo {author} {\bibfnamefont {S.~A.}\ \bibnamefont
  {Komrakova}}, \bibinfo {author} {\bibfnamefont {V.~N.}\ \bibnamefont
  {Sorokin}}, \bibinfo {author} {\bibfnamefont {P.~I.}\ \bibnamefont
  {Belobrov}}, \bibinfo {author} {\bibfnamefont {A.~N.}\ \bibnamefont
  {Smolyaninov}}, \ and\ \bibinfo {author} {\bibfnamefont {A.~V.}\ \bibnamefont
  {Akimov}},\ }\href {\doibase 10.1364/ome.7.004038} {\bibfield  {journal}
  {\bibinfo  {journal} {Opt. Mater. Express}\ }\textbf {\bibinfo {volume}
  {7}},\ \bibinfo {pages} {4038} (\bibinfo {year} {2017})}\BibitemShut
  {NoStop}%
\bibitem [{\citenamefont {Bradac}\ \emph {et~al.}(2009)\citenamefont {Bradac},
  \citenamefont {Gaebel}, \citenamefont {Naidoo}, \citenamefont {Rabeau},\ and\
  \citenamefont {Barnard}}]{Bradac2009}%
  \BibitemOpen
  \bibfield  {author} {\bibinfo {author} {\bibfnamefont {C.}~\bibnamefont
  {Bradac}}, \bibinfo {author} {\bibfnamefont {T.}~\bibnamefont {Gaebel}},
  \bibinfo {author} {\bibfnamefont {N.}~\bibnamefont {Naidoo}}, \bibinfo
  {author} {\bibfnamefont {J.~R.}\ \bibnamefont {Rabeau}}, \ and\ \bibinfo
  {author} {\bibfnamefont {A.~S.}\ \bibnamefont {Barnard}},\ }\href {\doibase
  10.1021/nl9017379} {\bibfield  {journal} {\bibinfo  {journal} {Nano Lett.}\
  }\textbf {\bibinfo {volume} {9}},\ \bibinfo {pages} {3555} (\bibinfo {year}
  {2009})}\BibitemShut {NoStop}%
\end{thebibliography}%
\end{document}